\documentclass[12pt]{article}
\usepackage{graphicx}
\usepackage{epsfig}
\usepackage{epstopdf}
\DeclareGraphicsExtensions{.pdf,.eps,.png,.jpg,.mps}

\def\lsim{\mathrel{\rlap{\lower3pt\hbox{\hskip0pt$\sim$}}
     \raise1pt\hbox{$<$}}}         
\def\gsim{\mathrel{\rlap{\lower4pt\hbox{\hskip1pt$\sim$}}
     \raise1pt\hbox{$>$}}}         

\usepackage{amsmath}
\usepackage{amsfonts}

\begin{document}
\begin{titlepage}

\centerline{\Large \bf Does the Universe have a Hard Drive?}
\medskip

\centerline{Zura Kakushadze$^\S$$^\dag$\footnote{\, Zura Kakushadze, Ph.D., is the President of Quantigic$^\circledR$ Solutions LLC, an Adjunct Professor at the University of Connecticut, and a Full Professor at Free University of Tbilisi. Email: \tt zura@quantigic.com}}
\bigskip

\centerline{\em $^\S$ Quantigic$^\circledR$ Solutions LLC}
\centerline{\em 1127 High Ridge Road \#135, Stamford, CT 06905\,\,\footnote{\, DISCLAIMER: This address is used by the corresponding author for no
purpose other than to indicate his professional affiliation as is customary in
publications. In particular, the contents of this paper
are not intended as an investment, legal, tax or any other such advice,
and in no way represent views of Quantigic® Solutions LLC,
the website \underline{www.quantigic.com} or any of their other affiliates.
}}
\centerline{\em $^\dag$ Free University of Tbilisi, Business School \& School of Physics}
\centerline{\em 240, David Agmashenebeli Alley, Tbilisi, 0159, Georgia}
\medskip
\centerline{(December 3, 2016)}

\bigskip
\medskip

\begin{abstract}
{}We discuss an apparent information paradox that arises in a materialist's description of the Universe if we assume that the Universe is 100\% quantum. We discuss possible ways out of the paradox, including that Laws of Nature are not purely deterministic, or that gravity is classical. Our observation of the paradox stems from an interdisciplinary thought process whereby the Universe can be viewed as a ``quantum computer". Our presentation is intentionally nontechnical to make it accessible to as wide a readership base as possible.
\end{abstract}
\medskip
\end{titlepage}

\newpage

{}Every time you start up your laptop, tablet or smart phone, its operating system\footnote{\, E.g., Linux, Windows, iOS, etc.} -- which sets the rules by which your device functions -- is loaded from a hard drive, where the operating system is stored, to its memory, where various processes run and computations are performed (for a schematic depiction, see Figure 1). Laws of Nature -- i.e., the laws of physics -- are analogous to an operating system by which the Universe functions. Thus, (almost)\footnote{\, There are things we still do not quite understand, e.g., the nature of dark matter, where the dark energy (a.k.a. the cosmological constant) comes from, how gravity fits in the quantum Universe (see below), etc. However, this does not affect the point we make herein.} everything we observe appears to be described by four fundamental forces: gravity, electromagnetism, weak and strong interactions.\footnote{\, The weak force is responsible for radioactive decay. The electromagnetic and weak interactions are unified into the electroweak force. The strong force binds quarks and gluons together inside protons, neutrons and nuclei comprised therefrom.} The latter three, together with the known subatomic particles,\footnote{\, These are leptons (e.g., electrons and neutrinos) and quarks. There is also the Higgs particle.} are described by the Standard Model of particle physics \cite{PDB}.\footnote{\, Which is sometimes also referred to as ``the theory of almost everything" \cite{Oerter}.}

{}Theoretically, the laws of physics are postulated. Then they are verified experimentally. These laws contain a nonzero amount of {\em information}, i.e., there is more than 0 bits of information encoded in them. Thus, if we, say, write down Einstein's celebrated mass-energy relation\footnote{\, $E$ is energy, $m$ is mass, and $c$ is the speed of light in the vacuum. Here we could have written down, e.g., Newton's second law $F = m~a$ ($F$ is force, $m$ is mass, and $a$ is acceleration). However, perhaps somewhat ironically, it appears that most people would recognize Einstein's $E = m~c^2$ \cite{Einstein} before Newton's $F = m~a$ \cite{Newton}.}
\begin{equation}
 E = m~c^2\,,
\end{equation}
this expression contains a nontrivial amount of information. So do all laws of physics we currently believe to describe Nature, including the Standard Model, Einstein's General Relativity, etc.\footnote{\, In the Standard Model one usually writes down its Lagrangian, which encodes classical propagation and interactions of various fundamental particles such as electrons, quarks, photons (quanta of light), etc., augmented with (the so-called quantum field theory) rules for computing quantum effects (via, e.g., path integral and Feynman diagrams \cite{Feynman}). For gravity we still do not have an experimentally verified quantum theory (string theory being a candidate therefor; see, e.g., \cite{String}); however, Einstein's General Relativity \cite{GR} appears to accurately describe classical gravity and its Lagrangian too encodes a nontrivial amount of information.} This compels us to ask the following question:

{}{\em How (where) does the Universe store the information encoded in the laws of physics?} This is by far not a rhetorical question. In a materialist's\footnote{\, That is, without resorting to a higher power of any kind. See, e.g., \cite{Sagan}.} description of Nature, any nontrivial amount of information -- including that encoded in the laws of physics -- must somehow be stored (be it in the form of matter or energy), just as an operating system is stored on your laptop's or smart phone's hard drive. From this (i.e., the materialist's) standpoint, it would be a copout -- unappealing both intellectually and scientifically -- to simply state to the effect that ``Laws of Nature are a property of the Universe". And this is where an apparent paradox arises...

{}On the fundamental level, our Universe is not deterministic but probabilistic -- the Universe is quantum.\footnote{\, This statement requires a qualification -- see below.} Assuming the Universe is 100\% quantum, i.e., all matter and interactions\footnote{\, Including gravity -- see below.} are described by a quantum theory, any information stored in the Universe cannot be purely deterministic but probabilistic. That is, the laws of physics apparently could not be deterministic. However, {\em our} description of the Universe via the laws of physics assumes that the latter are 100\% deterministic...

{}So, something's got to give. One possibility is that the laws of physics are not deterministic but probabilistic. This would amount to a major shift in the existing paradigm. Another possibility is that not all matter or interactions are quantum. Based on their precision tests, which incorporate quantum corrections, it appears to be a safe bet to assume that the electroweak and strong interactions are quantum. However, despite theoretical arguments\footnote{\, These are based on thought experiments such as those of \cite{Eppley} and \cite{Page}, whose critiques appear, e.g., in \cite{Mattingly} and \cite{Hawkins}, respectively. Thus, precisely due to the extreme weakness of gravity, to detect potential ``ill-effects" of coupling quantum matter to classical gravity (e.g., indefinite dissipation of energy by quantum matter via classical gravitational radiation), it would appear to require a detector so massive that it would be inside its own Schwarzschild radius \cite{Schwarzschield} (i.e., it would be a black hole), or detection might take much longer than the age of the Universe, etc. Here we will not delve into such controversy but simply note that, notwithstanding any (important) theoretical considerations, physics is an experimental science.} that gravity must also be quantum, currently there is no experimental evidence that gravity is in fact quantum, which stems from the extreme weakness of gravity (compared with other fundamental forces) at the microscopic level where quantum effects become relevant. So, perhaps one can consider a scenario where gravity is classical\footnote{\, Again, notwithstanding the aforesaid theoretical arguments to the contrary.} and the laws of physics are somehow encoded via gravity -- albeit here we do not hold ourselves out to understand how this would be realized in detail. Yet another apparent possibility would be to declare that the laws of physics are what they are, and that we do not need to understand how the information encoded in them is stored.\footnote{\, Here one can kick the can down the road by invoking the string landscape approach \cite{Landscape}, whereby, in a presumably unified string theory description, different looking universes -- including our Universe -- arise as different vacua, i.e., solutions to the string equations of motion. However, such a description itself encodes a nontrivial amount of information (e.g., in the form of the aforesaid string equations of motion), and assuming that string theory is quantum, we are still facing the same paradox, albeit perhaps on a deeper level. Furthermore, the many-worlds interpretation of quantum mechanics \cite{Everett} does not appear to alleviate the paradox.} However, this would appear to abandon the aforesaid materialist's standpoint -- one can then simply call the unquestionable and almighty laws of physics ``God" and be done with it...\footnote{\, And here we by no means suggest that this is not the right approach. After all, {\em arguendo}, this would appear to bode well with G\"odel's incompleteness theorems \cite{Godel}.}

{}The upshot is that we appear to have a paradox if we take the materialist's approach and assume that the Universe is 100\% quantum. The purpose of this note is to point out the paradox, not to resolve it. In this regard, we could simply conclude here. However, the paradox appears to run even deeper in the context of the Big Bang framework \cite{BigBang}, whereby our Universe was created about 14 billion years ago out of ``nothing" (vacuum). A priori there is no issue with creating matter and interactions out of ``nothing": matter carries positive (kinetic) energy, while gravity supplies negative (potential) energy, the total energy is zero and is therefore conserved when the Universe is created from ``nothing". However, the nonzero bits of information encoded in the laws of physics would have to be created out of ``nothing" within the Big Bang framework. Conceptually, this appears to be just as unappealing as the information loss paradox in the context of black hole evaporation via Hawking radiation (a quantum effect...) \cite{Hawking} -- ``structured" information (pure quantum state) encoded in the matter falling into a black hole is transformed into ``unstructured" information (mixed quantum state) carried by Hawking radiation (a.k.a. non-unitary evolution). Is creating structured information out of thin air any better conceptually than such information loss?

{}We do not know the answer(s) to the question(s) we raise herein. However, we believe raising them is warranted. Thinking about the laws of physics as encoding a nonzero amount of information is a very ``computer science" thing to do. In fact, we can think about the apparent paradox we describe in this note in the context of computer science. If we think about the Universe as a ``quantum computer" (cf. \cite{Lloyd}), we still need purely deterministic (classical) code -- i.e., the laws of physics -- to run it, which is how actual quantum computers work. The underlying physical processes in quantum computers are quantum, but the code is 100\% deterministic \cite{QC}. In this regard, finally, let us mention yet another possible way out of the aforesaid paradox: the simulation hypothesis \cite{Sim}, whereby our Universe is simply a computer simulation run by ``other beings". However, while the simulation hypothesis, {\em arguendo}, may not necessarily invalidate the scientific method (see, e.g., \cite{SimHypSci}), it does appear to kick the can down the road as we essentially have no way of describing the universe in which the ``other beings" live and run their curious simulations...

\subsection*{Acknowledgments}
{}I would like to thank Alberto Iglesias and Willie Yu for invaluable discussions and encouragement.

\bigskip
\bigskip
\bigskip
\begin{figure}[ht]
\centerline{\epsfxsize 4.truein \epsfysize 4.truein\epsfbox{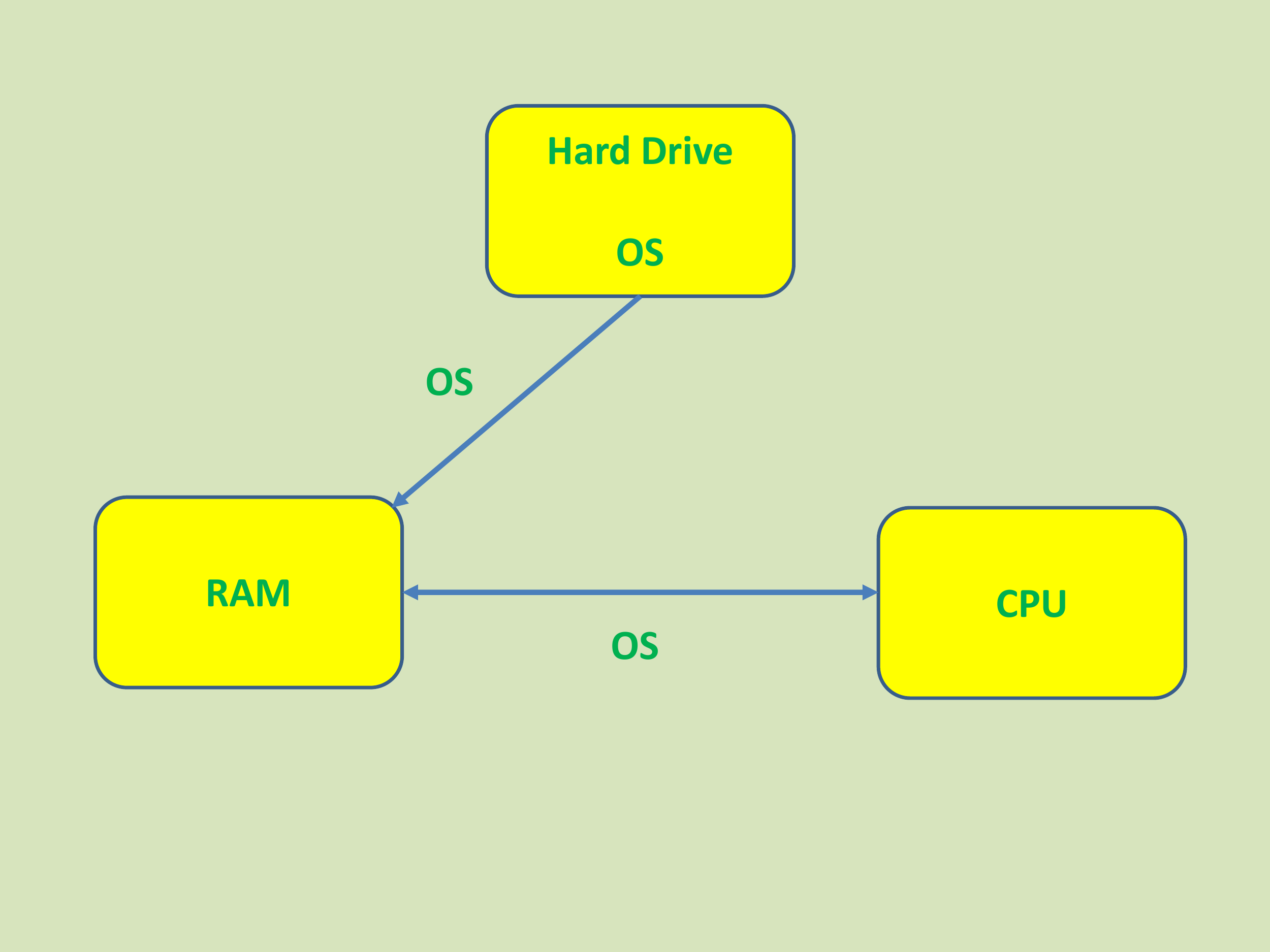}}
\noindent{\small {Figure 1. OS = Operating System. RAM = Random Access Memory. CPU = Central Processing Unit. RAM is the volatile memory space that stores the data directly accessed by CPU. OS is installed and stored on the Hard Drive (HD). The first software to run upon startup is BIOS (Basic Input/Output System), which loads OS into RAM from HD.}}
\end{figure}

\end{document}